\documentclass[12pt,preprint]{aastex}
\usepackage{natbib}
\usepackage{graphicx}

\slugcomment{Accepted for Publication in ApJ}

\begin{document}

\title{A Deep {\em Chandra} ACIS Study of NGC 4151. II. The Innermost
  Emission Line Region and Strong Evidence for Radio Jet--NLR Cloud
  Collision}

\author{Junfeng Wang\altaffilmark{1}, Giuseppina
  Fabbiano\altaffilmark{1}, Martin Elvis\altaffilmark{1}, Guido
  Risaliti\altaffilmark{1,2}, Carole G. Mundell\altaffilmark{3},
  Margarita Karovska\altaffilmark{1} and Andreas
  Zezas\altaffilmark{1,4}}

 \altaffiltext{1}{Harvard-Smithsonian Center for Astrophysics, 60 Garden St, Cambridge, MA 02138}
 \altaffiltext{2}{INAF-Arcetri Observatory, Largo E, Fermi 5, I-50125 Firenze, Italy}
 \altaffiltext{3}{Astrophysics Research Institute, Liverpool John Moores University, Birkenhead CH41 1LD, UK}
 \altaffiltext{4}{Physics Department, University of Crete, P.O. Box 2208, GR-710 03, Heraklion, Crete, Greece}


\email{juwang@cfa.harvard.edu}

\begin{abstract}

We have studied the X-ray emission within the inner $\sim$150 pc
radius of NGC 4151 by constructing high spatial resolution emission
line images of blended OVII, OVIII, and NeIX. These maps show extended
structures that are spatially correlated with the radio outflow and
optical [OIII] emission.  We find strong evidence for jet--gas cloud
interaction, including morphological correspondences with regions of
X-ray enhancement, peaks of near-infrared [FeII] emission, and optical
clouds. In these regions, moreover, we find evidence of elevated
NeIX/OVII ratios; the X-ray emission of these regions also exceeds
that expected from nuclear photoionization.  Spectral fitting reveals
the presence of a collisionally ionized component.  The thermal energy
of the hot gas suggests that $\ga 0.1\%$ of the estimated jet power is
deposited into the host interstellar medium through interaction
between the radio jet and the dense medium of the circum-nuclear
region.  We find possible pressure equilibrium between the
collisionally ionized hot gas and the photoionized line-emitting cool
clouds.  We also obtain constraints on the extended iron and silicon
fluorescent emission.  Both lines are spatially unresolved. The upper
limit on the contribution of an extended emission region to the Fe
K$\alpha$ emission is $\la 5\%$ of the total, in disagreement with a
previous claim that 65\% of the Fe K$\alpha$ emission originates in
the extended narrow line region.

\end{abstract}


\section{Introduction}

Multi-wavelength high spatial resolution observations of NGC 4151
\citep[e.g.,][]{Hutchings99,Yang01, Mundell03,Wang09_HRC, SB09} have
provided unique information on the continuum and emission line
morphology in the central few 100 pc \citep[$d\sim 13.3$ Mpc,
  $1\arcsec=65$ pc,][]{Mundell99}, contributing to our understanding
of the nature of the nuclear emission and its interaction with the
host galaxy (see \citealt{Ulrich00} for a review). In this paper we
investigate further these topics, using high quality deep {\em
  Chandra} ACIS observations of the nearby prototypical Seyfert 1
galaxy NGC 4151, which allow the morphological investigation of
spectral line emission in the inner circum-nuclear region ($r\sim 150$
pc in projection). In particular, we will address the radio jet--host
galaxy interaction and the nature of the Fe K$\alpha$ line emission.

High resolution radio surveys show broad existence of collimated radio
outflows or jets in Seyfert galaxies, albeit less prominent than those
of radio galaxies
\citep[e.g.,][]{Ulvestad84,Kukula95,Nagar99,Xanthopoulos10}. Ever
since the {\em Hubble Space Telescope} emission line imaging resolved
the NLR morphology in a number of Seyfert galaxies with linear radio
structures \citep{Capetti95,Falcke98,Cecil00}, it has been actively
debated whether the radio jet plays a competing role against nuclear
photoionization in the narrow-line region (NLR) ionization structure
\citep{Whittle88,Binette96, Dopita96, Bicknell98, Wilson99, Rossi00,
  Whittle04, Rosario07}.  {\em Chandra} X-ray observations of the
nuclear regions of nearby Seyfert galaxies appear to offer powerful
diagnostics for discriminating between photoionization by the nuclear
radiation or collisional ionization by a radio jet
\citep{Sako00,Young01}, although the importance of radio jets in
shaping NLR environments seems to differ case-by-case
\citep{Whittle05,Evans06,Gandhi06,Rosario10b,Rosario10a,Bianchi10}.

The nuclear region of NGC 4151 is known to host a two-sided,
$\sim$300~pc-long radio outflow along position angle (P.A.)
$\sim$77$^{\circ}$ \citep[e.g.,][]{Johnston82,Pedlar93,Mundell95}.
Within the central 100 pc of the galaxy, \citet{Mundell03} identified
a faint, highly collimated jet underlying discrete components that are
shock-like features associated with the jet--gas clouds interactions.
Although the distribution of the ionized gas and the kinematics in the
NLR imply that shock ionization originated from jet--cloud
interactions is unlikely to be the dominant source of ionizing photons
compared to the nucleus \citep{Kaiser00,Mundell03}, \citet{Mundell03}
find evidence that some NLR clouds are responsible for bending the
jet.  Near-IR emission-line mapping of the nuclear region by
\citet{SB09} also reveals enhancement of [FeII] emission, suggesting
the presence of shock heating in addition to nuclear photoionization
\citep{SB09}.

In our previous work using the {\em Chandra} HRC image of NGC 4151,
\citet{Wang09_HRC} found an excess of X-ray emission at these
jet--cloud interaction locations.  However, no spectral information
could be obtained because of the very limited energy resolution of the
HRC.  Previous {\em Chandra} ACIS spectral imaging data of the NGC
4151 nuclear region \citep{Ogle00,Yang01} were limited by both heavy
pile-up\footnote{See the Chandra ABC Guide to Pileup available at
  \url{http://cxc.harvard.edu/ciao/download/doc/pileup\_abc.pdf}} and
sensitivity, and did not allow detailed high spatial resolution
comparison to the radio jet and emission line gas, nor spectral
studies of X-ray emission from spatially resolved features. Our 200 ks
ACIS-S observation of NGC 4151 \citep[PI Fabbiano; see][and Wang et
  al.\ 2011 for previous work on this data set]{Wang10_EXT} allows us
to pursue these investigations, by comparing the morphology of the
NeIX and OVII with that of the radio jet.

Moreover, with these data we can address the outstanding controversy
on the spatial properties of the 6.4~keV Fe K$\alpha$ emission of NGC
4151. In Seyfert galaxies, this emission line is expected to originate
from cold matter near the nucleus ($<1$~pc), either the obscuring
torus \citep[e.g.,][]{Krolik87} or an accretion disk
\citep[e.g.,][]{George91, Reynolds97}. These model predictions,
however, conflicted with the finding of \citet{Ogle00}, who reported
spatially resolved narrow iron K$\alpha$ line emission in the {\em
  Chandra} HETG observation, concluding that $65\% \pm9\%$ of this
emission originates in the ENLR at distance up to 6\arcsec\/ from the
nucleus ($\sim$400 pc across). This conclusion has been more recently
contested by the {\em XMM}-Newton based work of \citet{Schurch03}.

\section{Observations and Data Analysis}

The details of our ACIS observations and data reduction are described
in \citet[][hereafter Paper I]{Wang11}.  Briefly, NGC 4151 was
observed by {\em Chandra} for a total of 180 ks (after screening for
high background intervals) with the spectroscopic array of the
Advanced CCD Imaging Spectrometer \citep[ACIS-S;][]{Garmire03} in 1/8
sub-array mode during March 27-29, 2008.  The data were reprocessed
following standard procedures, using CIAO (Version 4.2) with the CALDB
4.2.1 provided by the {\em Chandra} X-ray Center (CXC).  Subpixel
event repositioning (``static'' method in Li et al. 2004, using the
corner split and 2-pixel split events) and subpixel binning techniques
were applied to the ACIS images to improve the spatial resolution.

During our previous work \citep{Wang10_NUC} and Paper I, we became
well aware of the complexity in data analysis caused by the bright
nuclear emission of NGC 4151.  First, photon pile-up is present in the
nuclear region even with the reduced frame-time of our observation.
We established that for the soft X-ray emission, pile-up is moderate
($<10\%$) at $r > 1 \arcsec$.  Second, although the {\em Chandra}'s
PSF is highly centrally peaked, the contamination from the scattered
nuclear emission to the extended emission is not negligible due to the
brightness of the nucleus and the broader PSF in the higher energy
range.  We performed {\em Chandra} PSF simulations that provide an
estimate of the expected contamination from the nuclear emission in an
extended feature.  Detailed explanation of the analysis leading to
these conclusions is presented in Paper I.  We have taken into account
this information in the following analysis.

\subsection{Soft X-ray Emission Line Images}

The 2--7~keV emission is dominated by the unresolved nucleus
\citep{Yang01,Wang09_HRC}. Since we are interested in the extended
X-ray emission, we used the energy range below 2 keV where this
emission is prominent (see Paper I).  To investigate the general
spectral dependency of the morphology of this soft X-ray emission,
images in three spectral bands below 2 keV were extracted from the
merged data: 0.3--0.7 keV (``soft band''), 0.7--1.0 keV (``medium
band''), and 1--2 keV (``hard band'').  Following \citet{Wang09_1365},
exposure maps were created for the individual bands to obtain
exposure-corrected flux images.

\begin{figure}
\epsscale{0.6}
\plotone{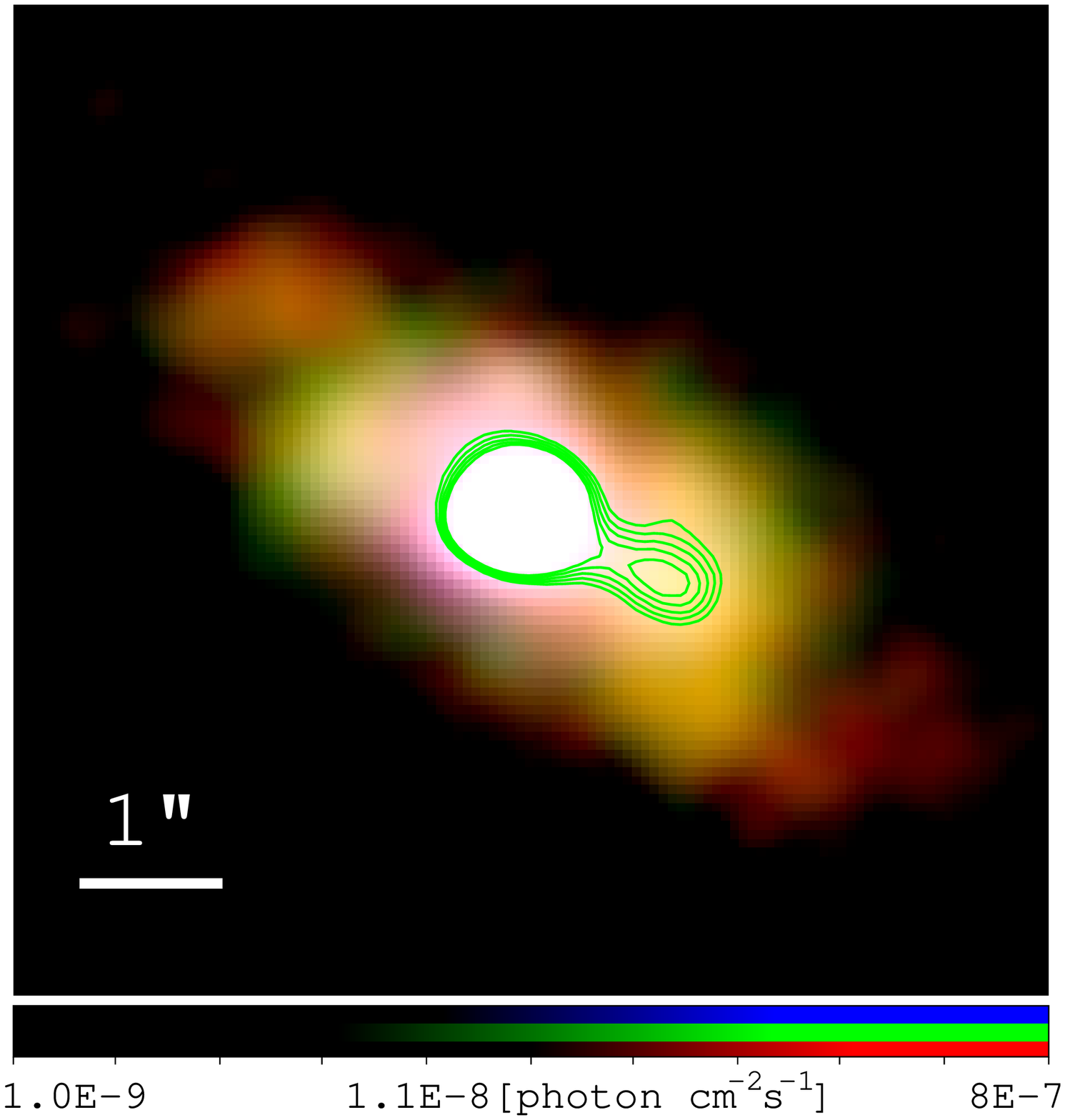}
\epsscale{1.0}
\caption{The tri-color composite image of the central $7\arcsec \times
  7\arcsec$ (450~pc-across) circumnuclear region of NGC 4151, where
  the soft (0.3--0.7 keV), medium (0.7--1 keV), and hard band (1--2
  keV) images have been smoothed with a $FWHM=0.3\arcsec$ Gaussian kernel and
  shown in red, green, and blue, respectively.  The contours are of
  the medium band and highlight a linear feature spatially coincident
  with the radio jet.  The pixel scale is $0.0625\arcsec$ per pixel,
  1/8 the native ACIS pixel.
\label{3color}}
\end{figure}

Figure~\ref{3color} presents the resulting false color composite image
of the central $7\arcsec \times 7\arcsec$ (450~pc on a side) region of
NGC 4151, where the soft, medium, and hard band images are shown in
red, green, and blue, respectively. The images in the three band have
been smoothed with a $FWHM=0.3\arcsec$ Gaussian kernel.  It clearly
shows bright structured soft X-ray emission along the northeast (NE)
-- southwest (SW) direction, which is also the direction of the radio
jet.  In particular, the medium band image shows a jet-like
$\sim$2\arcsec\/ linear extension (contours in
Figure~\ref{3color}). This medium band is dominated by the NeIX
emission (see below).  In the following we concentrate on the analysis
of this inner 2\arcsec-radius circum-nuclear region.

High spectral resolution grating spectra of NGC 4151
\citep{Ogle00,Schurch04,Armentrout07} have shown that the emission in
these soft spectral bands is almost entirely dominated by lines.
Although the spectral resolution of ACIS CCD cannot provide unique
identifications of the strongest soft X-ray emission lines ($<$2~keV)
seen in the nuclear spectrum \citep{Wang10_NUC}, we can identify the
dominant transitions guided by the HETG observations \citep{Ogle00}.
Most notably the blended lines appear as three strong lines in the
ACIS spectrum, approximately centered at 0.57~keV, 0.68~keV, 0.91~keV
(c.f. Figure 4 in Yang et al.\ 2001; Figure 3 in
\citealt{Wang10_NUC}).  We then extracted the X-ray emission in three
narrow energy intervals (0.53--0.63 keV, blended OVII f,i,r;
0.63--0.73 keV, blended OVIII Ly$\alpha$ and OVII RRC; 0.85--0.95 keV,
blended NeIX f,i,r) to create line strength images, highlighting
regions of these prominent emission lines.  This is a reasonable
approach since the line emission dominates over the weak underlying
continuum \citep{Ogle00,Schurch04} in these narrow bands and reveal
substructures that are not obvious in the broadband images
\citep[e.g.,][]{Bianchi10,GM10}.

\begin{figure}
\includegraphics[scale=0.9]{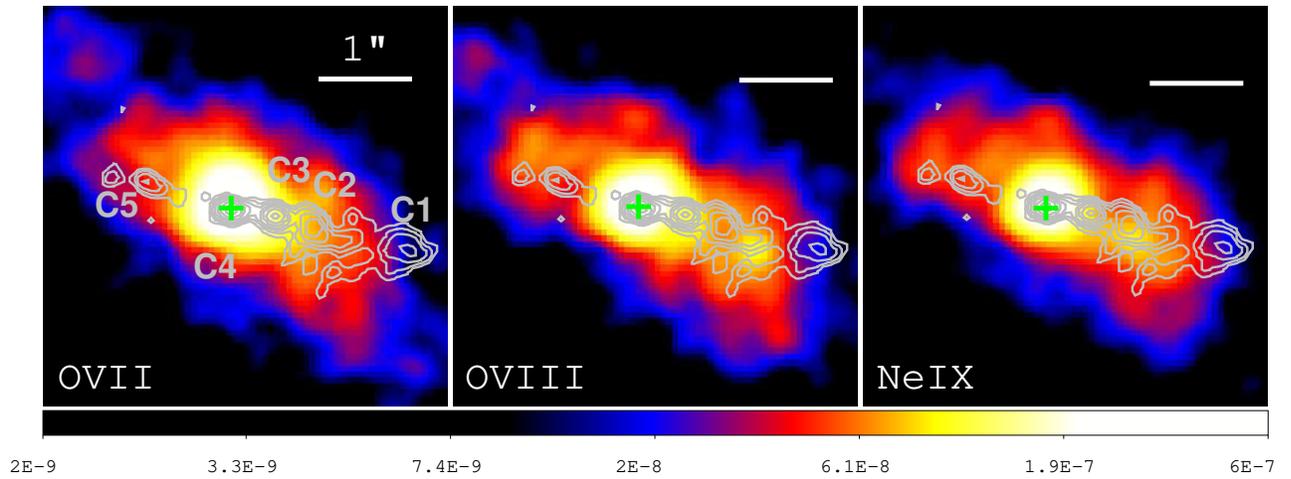}
\caption{The emission-line structure of the hot ISM in the central
  $\sim$250 pc of NGC 4151 (a) OVII; (b) OVIII+OVII RRC; (c) NeIX.
  The nucleus position is indicated with a cross.  The contours
  outline the radio outflow in the 1.4 GHz MERLIN map (Mundell et
  al.\ 1995). The ACIS images have been rebinned to $0.0625\arcsec$
  per pixel, and smoothed with a $FWHM=0.3\arcsec$ Gaussian kernel.
\label{lines}}
\end{figure}

\begin{figure}
\includegraphics[scale=0.9]{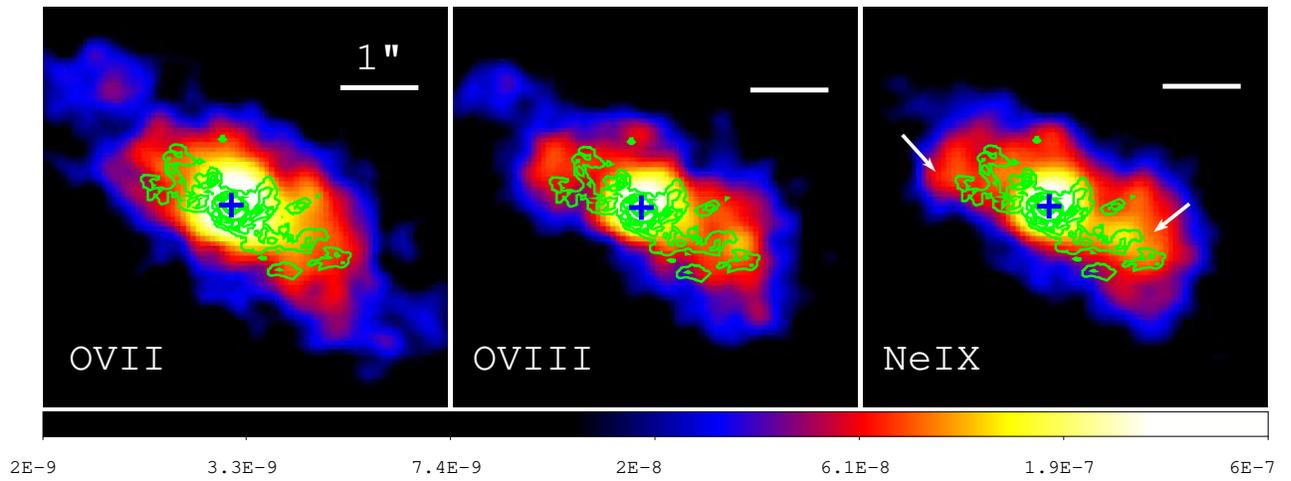}
\caption{Same emission-line maps of the central $\sim$250 pc of NGC
  4151 as in Figure~\ref{lines}, but with contours showing the
  $HST$/FOC [OIII]$\lambda 5007$ emission line clouds (Winge et
  al.\ 1997). (a) OVII; (b) OVIII+OVII RRC; (c) NeIX. The ACIS images
  have been rebinned to $0.0625\arcsec$ per pixel, and smoothed with a
  $FWHM=0.3\arcsec$ Gaussian kernel.
\label{o3}}
\end{figure}

The resulting images are shown in Figures~\ref{lines} and~\ref{o3},
showing the X-ray emission-line structure close to the nucleus (the
central $r=2\arcsec$, $\sim$130 pc) in the context of the radio
outflow and the optical emission line clouds.  The alignment between
these images is done using the peak of X-ray emission, [OIII]
emission, and the radio core.  Since the astrometry of our {\em
  Chandra} image is accurate to $0.3\arcsec$, our interpretation of
these features is not affected by significant alignment uncertainties.
The position of the nucleus is indicated with a cross.  The OVII,
OVIII, and NeIX emission line images all show extended morphology and
some structures, closely following the P.A. of the large scale
extended NLR traced by the [OIII] emission
\citep[e.g.,][]{Winge97,Kaiser00}.

\begin{figure}
\epsscale{1.0}
\plotone{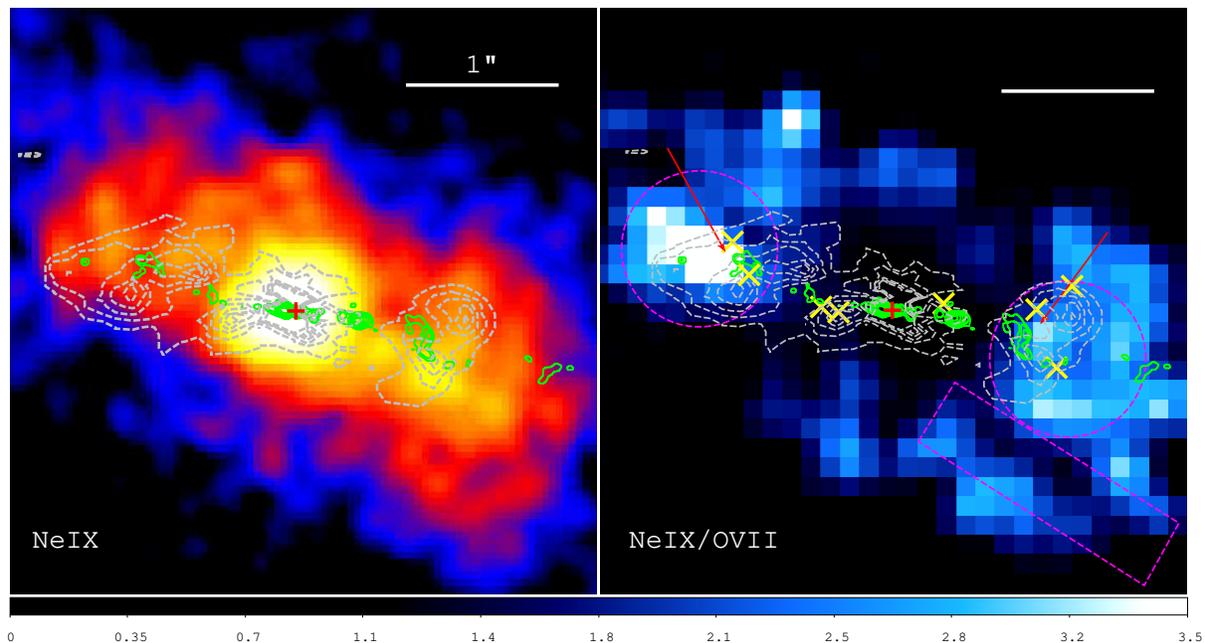}
\caption{(a) A close view of the NeIX emission line image overlaid
  with the contours of VLBA+VLA radio jet (green lines, Mundell et
  al. 2003) and NIR [FeII] (black dashed lines; \citet{SB09}). Note
  how the NeIX line emission peaks appears to align with the general
  direction of the linear radio outflow.  (b) Ratio image between the
  NeIX band and the OVII band images. Contours are the same as in (a).
  The arrows (red) indicate the knots of high NeIX/OVII ratios, and
  the circles (magenta) mark the spectra extraction regions. The box
  (magenta) marks the extraction region for the low NeIX/OVII ratio
  region. The crosses (yellow) mark the locations of the $HST$ clouds
  with high velocity dispersion (Kaiser et al.\ 2000; see also Figure
  7 in Mundell et al.\ 2003).
\label{jet}}
\end{figure}

In particular, the linear feature seen in the medium-band is clearly
present in the NeIX image.  There appears to be two X-ray enhancements
bracketed by the optical clouds (indicated by arrows in
Figure~\ref{o3}) that are close to the radio knot features C2 and C5
\citep{Mundell95}.  These ``hot spots'' are better visualized in the
NeIX/OVII ratio image, shown in Figure~\ref{jet} together with the
VLBA jet \citep{Mundell03} and the NIR [FeII] line emission
\citep{SB09}.  The locations where the jet appears to be intercepted
by the optical clouds and where prominent [FeII] emission arise, are
characterised by a NeIX/OVII ratio of $2.8\pm 0.2$, which is
significantly ($\sim 6\sigma$) higher than the ratio in the
surrounding regions ($1.2 \pm 0.2$).  We also examined the optical
clouds with high velocity dispersion, possibly associated with
jet-cloud impact (marked as crosses in Figure~\ref{jet}; see Figure~7
in Mundell et al.\ 2003), and find an elevated NeIX/OVII ratio of
$2.1\pm 0.4$ (with marginal significance).  The eastern NeIX/OVII hot
spot appears to be associated with clear interactions between the jet
and the gas cloud, where shock heating becomes important.  We note
that the observed line flux of OVII is more suppressed than NeIX in
the presence of the Galactic absorption column ($N_H=2\times 10^{20}$
cm$^{-2}$), and XSPEC \citep{Arnaud96} simulations show the absorption
corrected ratio may systematically decrease by 10\%.  Thus the
presence of NeIX/OVII hot spots is not affected, unless there is
significant differential absorption in the region.

\subsection{Spectral Analysis of the Soft Emission}\label{sec-large_scale}

We extracted X-ray spectra from both the high NeIX/OVII emission line
ratio regions and the surrounding low ratio region
(Figure~\ref{jet}b), and fit them jointly (with the same model
components but normalizations set free) for comparison using XSPEC
(Version 12.5; Arnaud 1996). Spectra and instrument responses were
generated using CIAO tool {\em
  specextract}\footnote{\url{http://cxc.harvard.edu/ciao/threads/specextract/}}.
The background spectrum is taken from a nearby source-free region on
the same CCD node.  Spectra were grouped to have a minimum of 20
counts per energy bin to allow for $\chi^2$ fitting.  The contribution
from the bright nuclear emission cannot be neglected here; to have a
self-consistent model, we included a fraction of the scattered
emission predicted from the PSF simulation of the point-like NGC 4151
nucleus.

We have made use of the {\tt Cloudy} photoionization modeling code,
last described by \citet{Ferland98}, to model the soft X-ray
emission. Using {\tt Cloudy} version C08.00\footnote{Available at
  \url{http://www.nublado.org/wiki/DownLoad}}, which enables a {\tt
  Cloudy}/XSPEC interface \citep{Porter06}, we attempted to produce
the soft part of the X-ray spectrum assuming an open plane-parallel
geometry (``slab'').  The dimensionless ionization parameter
\citep{Osterbrock06} is defined as $U=Q/(4\pi r^2cn_H)$, where $n_H$
is the hydrogen number density, $r$ is the distance to the inner face
of the model slab, $c$ is the speed of light, and
$Q=\int_{13.6eV}^{\infty}{L_{\nu}/h{\nu}}$ is the emitting rate of
hydrogen ionizing photons (s$^{-1}$) by the ionizing source.  It was
previously noted that the spectral fit to the soft X-ray continuum of
the nucleus is quite uncertain because of heavy absorption
\citep{Armentrout07}, therefore we adopted the broken power-law form
in \citet{Kraemer05} for the AGN continuum ($Gamma=2.3$ for the energy
range between 13.6 eV and 0.5 keV, and $Gamma=1.5$ for $E>0.5$~keV).
Normalizing to the observed X-ray luminosity of the nucleus
\citep{Wang10_NUC}, we obtain $Q\sim 7\times 10^{53}$ (photons
s$^{-1}$). We varied $U$ and the column density $N_H$ of the model
slab to create spectral model grids, which were fed to XSPEC.

\begin{figure}
\includegraphics[scale=.50,angle=-90]{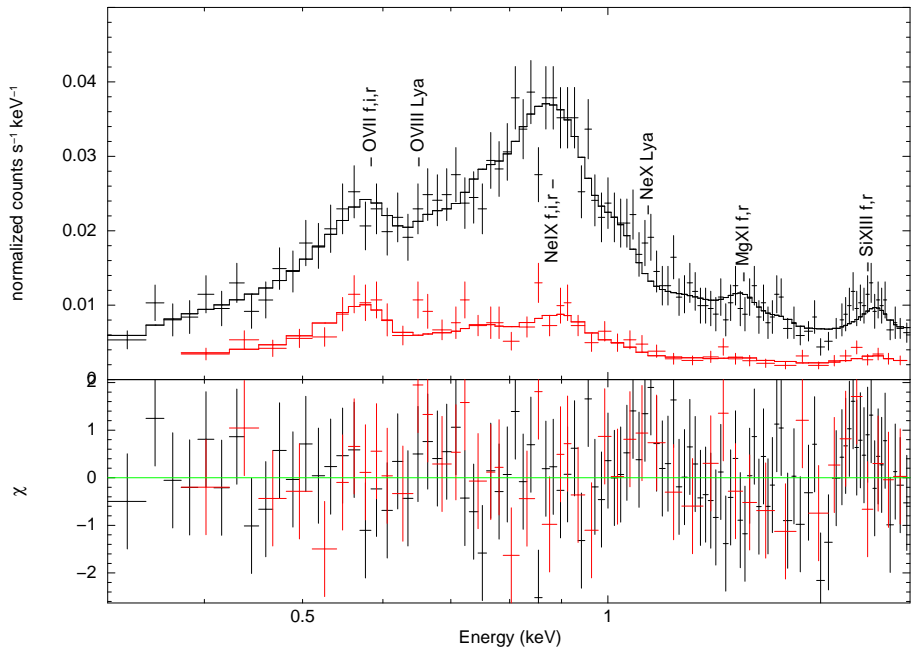}
\caption{Spectral fit of the X-ray emission in hot spot (black) and
  surrounding regions (red). Likely identifications of the blended
  emission lines are labeled, based on the HETG line list in
  \citet{Ogle00}. See text and Table~1 for model details.
\label{2reg}}
\end{figure}

We quickly find that, using a single photoionized component alone
cannot reproduce the observed soft X-ray emission in NGC 4151. With a
high photoionization component ($\log U \sim 1.9$), we were able to
produce the hydrogen-like neon and oxygen lines emission (e.g., OVIII,
NeX), which are unambiguously present in all the grating spectra
\citep[e.g.,][]{Ogle00, Schurch04, Armentrout07}.  However, the
helium-like transitions, in particular the OVII and NeIX emission line
features, are not fitted acceptably.  Adding a second lower ionization
photoionized component ($\log U \sim 0$), as detected in
\citet{Armentrout07}, we were able to obtain a significantly improved
fit ($\Delta \chi=200$), except for residuals in the 0.7--1.1 keV
range that are particularly strong for the hot spot spectrum.  Adding
additional photoionized components to the model does not further
improve the fit when an F-test is performed.  Allowing different
absorption columns for the models also does not give a better fit
(reduced $\chi^2_{\nu} = 1.7$), indicating that the high NeIX/OVII
ratio is not due to higher obscuration.  The absorption column
required by the fits is consistent with the low Galactic extinction
towards the NLR region found in \citet{Crenshaw05}.  Instead, these
residuals disappear when a thermal emission component
\citep[$APEC$;][]{APEC01} with a temperature of $kT=0.58\pm 0.05$~keV
is added to the model, suggesting the presence of collisionally
ionized Fe L emission.  The spectral fitting results are summarized in
Table~1 and shown in Figure~\ref{2reg}.  The absorption column is
consistent with the low Galactic extinction towards the NLR region
found in \citet{Crenshaw05}.

Lastly, we note that fitting the data with only combinations of
absorbed optically-thin thermal emission ($APEC$) gave strong
residuals of emission line features (reduced $\chi^2_{\nu}\gg 1$),
even when the abundance $Z$ is allowed to vary.

\subsection{[OIII]/Soft X-ray Ratios in the Substructures}

The subpixel resolution ACIS image also allows a similar comparison of
the X-ray emission with the optical NLR clouds resolved in the $HST$
image.  Taking advantage of the spectral capability of ACIS, we
measured the 0.3--2 keV counts for the same [OIII] clouds identified
in \citet{Wang10_NUC}, and derived their 0.5--2 keV X-ray fluxes using
the spectral model above.  To subtract the nuclear contribution, we
used a local region at the same radii but with different azimuthal
angles than the clouds.  The results are listed in Table~2.
Figure~\ref{ratio} shows these [OIII] to soft X-ray ratios, comparing
the measurements with HRC results in \citet{Wang10_NUC} and
photoionization models in \citet{Bianchi06}.

The [OIII]/X-ray ratios at the two X-ray hot spot locations
(corresponding to radio knots C2 and C5 in \citealt{Mundell95}) were
also measured, and found to be uniformly low, $\sim$3.  This was
previously noted in \citet{Wang09_HRC}, and implies enhanced X-ray
emission compared to other clouds under nuclear photoionization, which
is likely associated with the jet--cloud interaction and consistent
with their association with high NeIX/OVII ratio regions.

\subsection{Spatial Morphology of the Fluorescent Line Emission}

The 6.4~keV fluorescent iron K$\alpha$ emission of neutral or mildly
ionized cold material is readily visible in the ACIS spectra of the
nucleus \citep{Wang10_NUC} and the extended region
(Figure~\ref{FeK1}a).  Another fluorescent feature, SiI K$\alpha$
emission at 1.74~keV, is also clearly present in the extended emission
(Figure~\ref{FeK1}b).  Both spectra of extended emission were
extracted from a $2 \arcsec <r< 30\arcsec$ annular region centered on
the nucleus. Is this ``extended'' fluorescent line emission truly
originating from outside the nuclear region, or is it the result of
the PSF wing spreading out point-like nuclear features?

To determine the spatial distribution of the Fe K$\alpha$ line
emitting material, we followed the procedure in \citet{SW01}: a
continuum image, taken to be the average of the counts in the 5.9--6.2
keV and 6.5--6.8 keV bands, was subtracted from the image extracted in
the 6.2--6.5 keV band, resulting in a ``pure'' FeK$\alpha$ line image,
shown in Figure~\ref{FeK2}a.  The Fe K$\alpha$ emission appears mostly
circular in the $r\leq 2\arcsec$, but shows faint extended emission
towards the north and west.

The Si K$\alpha$ line is blended with the MgXII Ly$\beta$ and SiXIII
$f,r$ emission \citep{Ogle00}, thus the above continuum subtraction is
not applied.  We extracted emission in the 1.7--1.8 keV range to
obtain the Si K$\alpha$ emission map (Figure~\ref{FeK2}b).  It is
mostly concentrated within a $r\sim 2\arcsec$ circle, with a slight
elongation along the optical bi-cone direction (northeast--southwest).

However, at such energies the wings of the energy-dependent telescope
PSF spread a small but non-negligible amount of the nuclear photons to
large radii.  Thus the presence of ``extended'' Fe K$\alpha$ emission
or Si K$\alpha$ emission in the $r>1\arcsec$ region may in fact come
from the PSF-scattered unresolved nuclear emission.

To investigate this point, we further compared the observed radial
profile of the Fe K$\alpha$ emission band (6.2--6.5 keV) and the
radial profile in the same band for a point-like nucleus simulated
with MARX\footnote{Available at \url{http://space.mit.edu/CXC/MARX/}}
\citep[see][]{Wang10_NUC}, shown in Figure~\ref{FeK}a.  A similar
comparison for the Si K$\alpha$ (1.7--1.8 keV) is shown in
Figure~\ref{FeK}b.  The extracted observed profiles shown in
Figure~\ref{FeK} agree well with the PSF profiles, except for the
inner $r<2\arcsec$ where the simulation overpredicted the observed
emission due to heavy pile up in the PSF core.  

The simulated PSFs indicate that the $r\leq 2\arcsec$ region contains
89\% of the point source emission in the 6.2--6.5 keV band and 94\% of
the point source emission in the 1.7--1.8 keV band.  Approximately 6\%
of the nuclear Fe K$\alpha$ emission is expected to be spread within
the $2\arcsec \la r \la 6\arcsec$ extended regions, the extent of Fe
K$\alpha$ seen in Figure~\ref{FeK1}a.  Taking into account the
contribution from the PSF wings, the remaining extended Fe K$\alpha$
emission is $5\pm 1\%$ of the total Fe K$\alpha$ emission.  This
should be considered an upper limit for the truly extended emission,
because the nuclear Fe K$\alpha$ emission is likely to be
underestimated due to pile up.  Similarly, we estimated that the
extended Si K$\alpha$ emission is $\la 4\pm 1\%$ of the total Si
K$\alpha$ emission.  As pointed out by the referee, the measured
extended emission could easily disappear with a slightly broader
simulated PSF given that the DitherBlur parameter in the MARX
simulation is not known a priori.

Therefore, we conclude that although there is faint extended emission
seen in the Fe and Si fluorescent line images, at most 5\% of the
observed Fe K$\alpha$ emission and 4\% of the Si K$\alpha$ emission
can be truly spatially extended; most of these emissions are
consistent with the PSF scattering of a strong nuclear component.

\section{Discussion}

\subsection{Jet--Cloud Interaction in a Photoionized Outflow}

\begin{figure}
\plotone{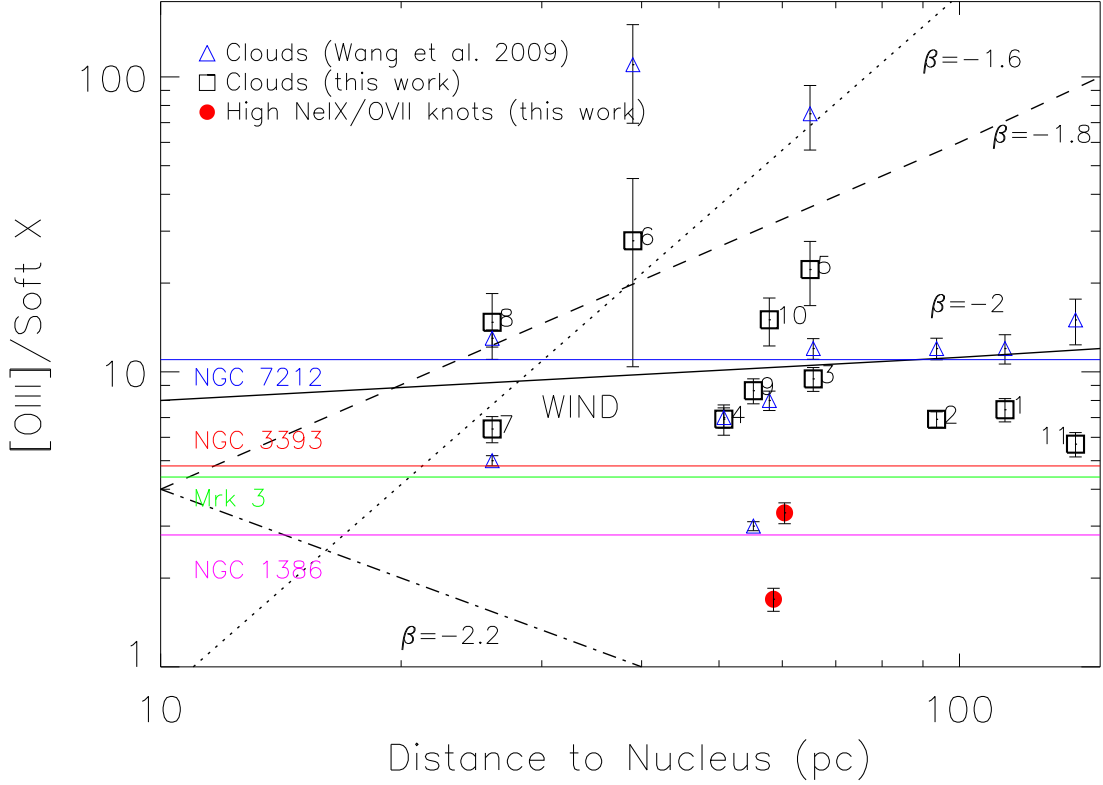}
\caption{The [OIII] to soft X-ray ratio as a function of the cloud's
  distance to the nucleus.  The triangles are the measurements in Wang
  et al.\ (2009), and the squares are from this work.  The filled
  (red) circles represent the two high NeIX/OVII ratio regions
  (Figure~\ref{jet}).  The blue, red, green, and magenta lines
  indicate the [OIII]/X-ray ratios for NGC 7212, NGC 3393, Mrk 3, and
  NGC 1386 (Bianchi et al. 2006), respectively.  The dotted, dashed,
  solid, and dot-dashed lines are the {\tt Cloudy} model predicted
  values from Bianchi et al. (2006) for different radial density
  profiles: $n_e\propto r^{\beta}$ where $\beta=-1.6$, $-1.8$, $-2$,
  and $-2.2$, respectively.
\label{ratio}}
\end{figure}

\begin{figure}
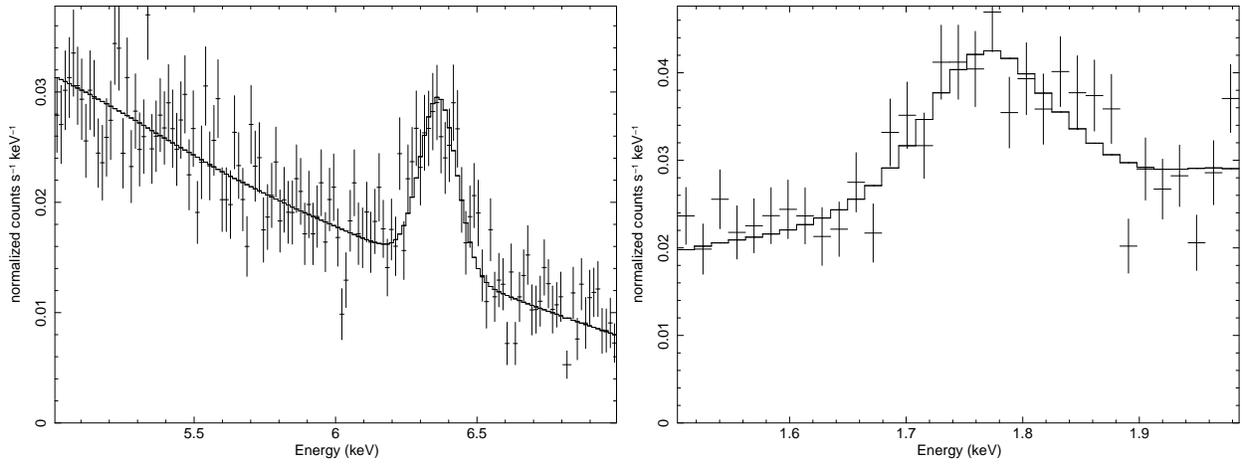

\includegraphics[scale=.35,angle=-90]{f7a.eps}
\includegraphics[scale=.35,angle=-90]{f7b.eps}
\caption{(a) The spectrum of the extended emission showing presence of
  Fe K$\alpha$ line.  (b) Same spectrum but showing the Si K$\alpha$
  line part.  The underlying solid line is the NGC 4151 nuclear
  spectrum (modeled with a power law plus a gaussian), scaled to 6\%
  and 3\% of the nuclear flux, respectively.
\label{FeK1}}
\end{figure}

\begin{figure}
\epsscale{0.5}
\plotone{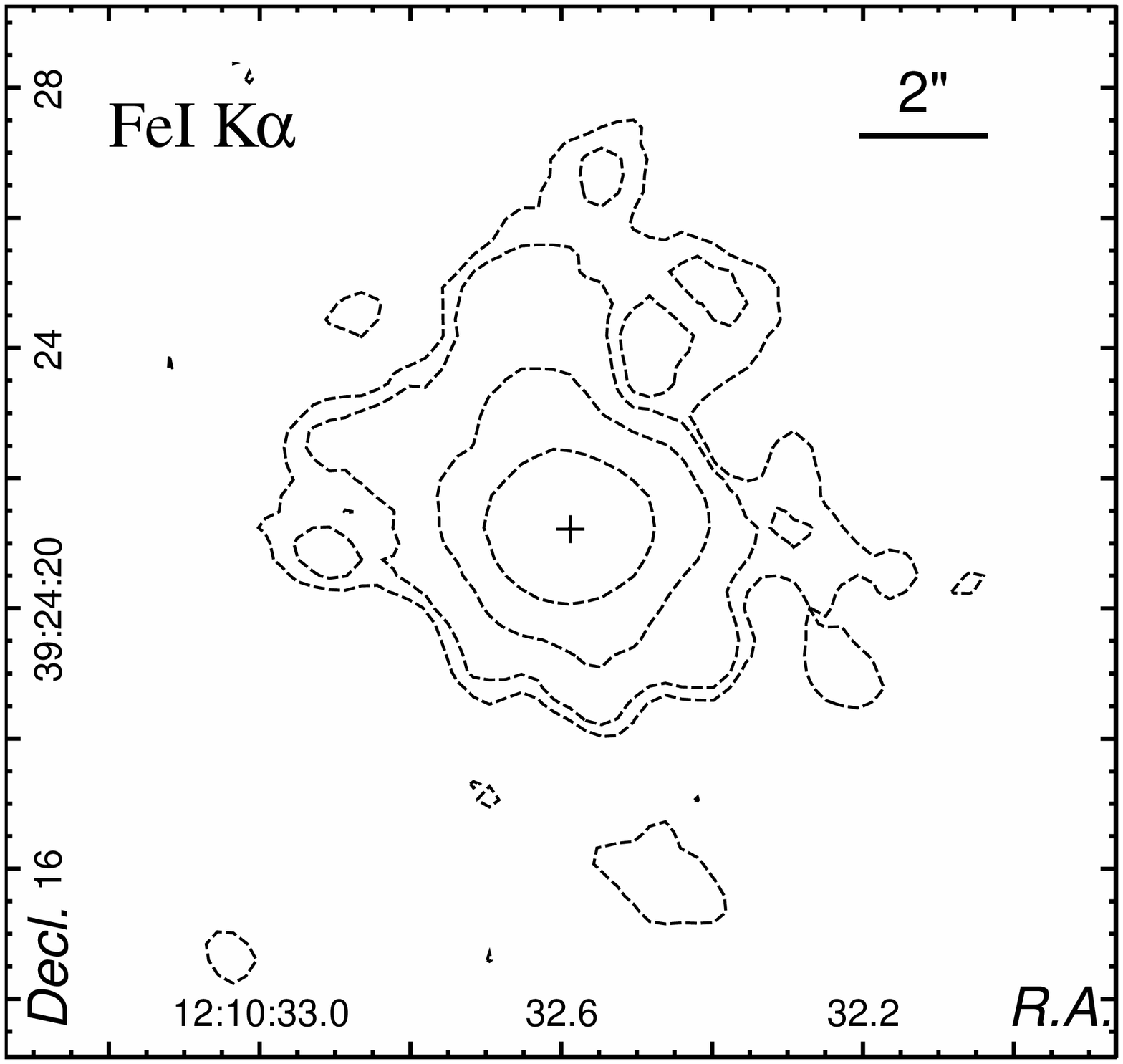}
\plotone{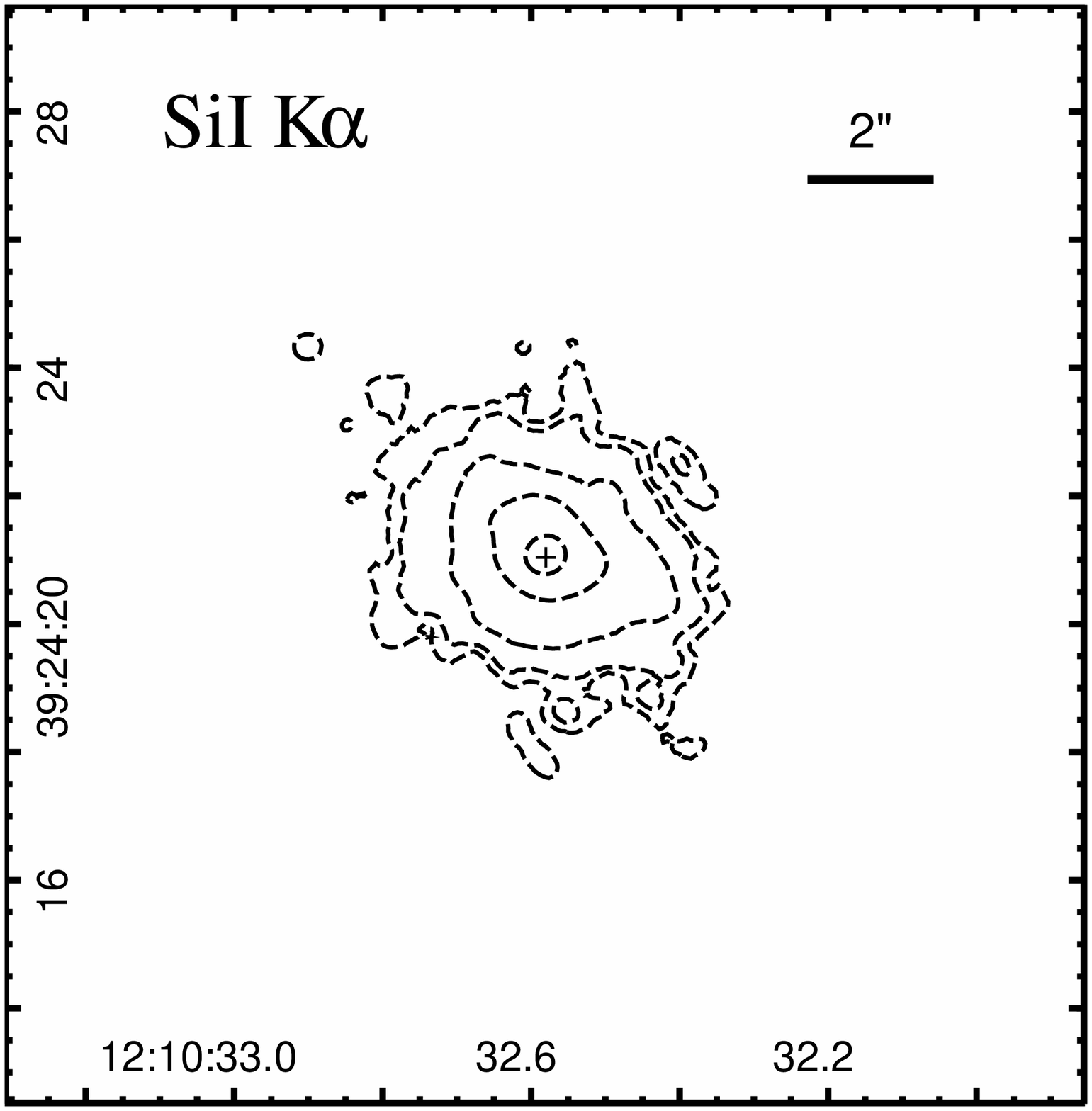}
\caption{(a) The contour map of continuum-subtracted Fe K$\alpha$ line
  emission of the central $15\arcsec \times 15\arcsec$ region in NGC
  4151. (b) The contour map of SiI K$\alpha$ line emission of the same
  region, extracted between 1.7 keV and 1.8 keV.  The cross marks
  position of the nucleus.  The contour levels are of logarithmic
  scale, between $3\times 10^{-10}$ to $4\times 10^{-8}$ photons
  cm$^{-2}$ s$^{-1}$.
\label{FeK2}}
\end{figure}

\begin{figure}
\epsscale{0.7}
\plotone{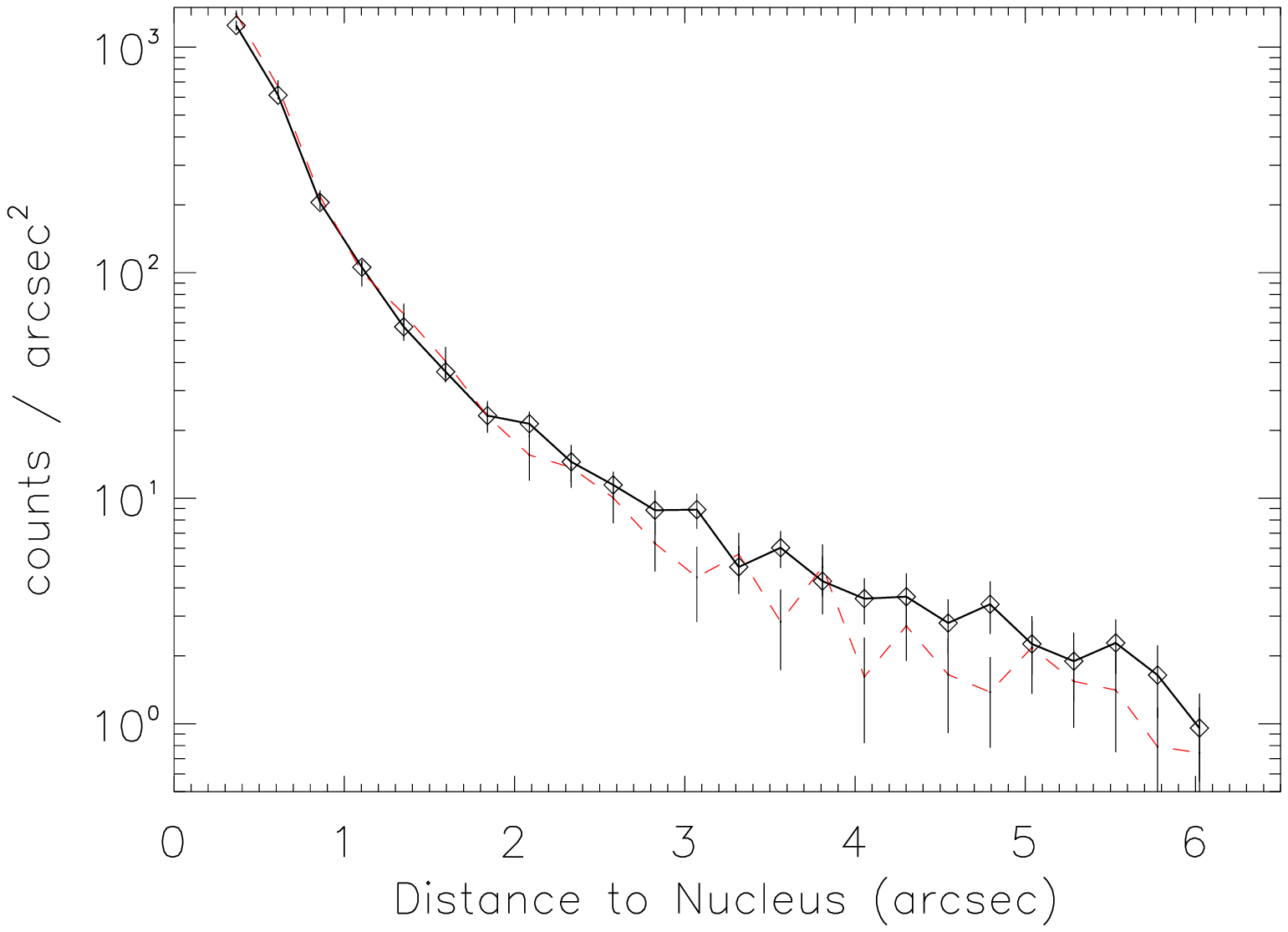}
\plotone{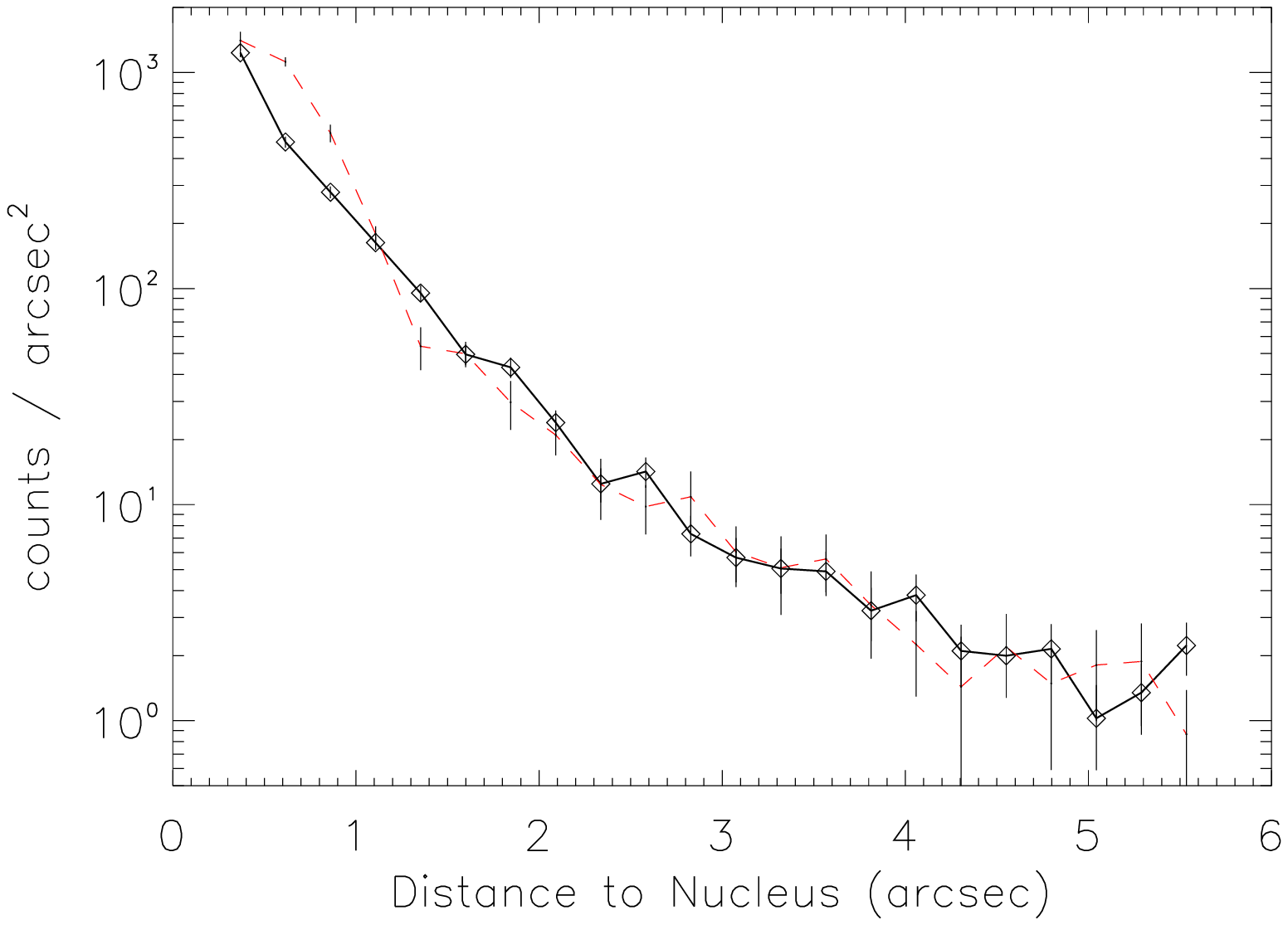}
\caption{(a) Comparison between the radial profile of simulated PSF in
  the 6.2--6.5 keV band (red dashed line) and the observed narrow-band
  Fe K$\alpha$ line emission (black solid line). (b) The same
  comparison but for the 1.7--1.8 keV Si K$\alpha$ line emission band.
\label{FeK}}
\end{figure}

Both our previous work using HRC image of NGC 4151 nuclear region
\citep{Wang09_HRC} and this work using ACIS find that most of the NLR
clouds in the central 150~pc radius region of NGC 4151 have a
relatively constant [OIII] to soft X-ray ratio ($\sim$10;
Figure~\ref{ratio} and Table~2), no matter the distance of the clouds
from the nucleus.  This ratio indicates a uniform ionization parameter
and a density decreasing as $r^{-2}$.  This may be consistent with a
photoionized wind scenario (e.g., Bianchi et al.\ 2006) as the NLR
clouds are outflowing (e.g., Kaiser et al.\ 2000).  The [OIII]/X-ray
ratios at the locations of jet-cloud collision are lower ($\sim$3),
leading to the suggestion of enhanced X-ray emission in these regions
due to shock heating in addition to photoionization.

This conclusion is supported by the results of multiwavelength imaging
studies.  The high spatial resolution emission line images of OVII,
OVIII, and NeIX emission show extended structures, in particular two
``hot spots'' in the NeIX image, which are close to the radio knot
features (Figures~\ref{lines} and~\ref{o3}).  At these locations
(1\arcsec\/ from the nucleus), \citet{Mundell03} noted that the
morphology of the highly collimated radio outflows appears more
disturbed, suggesting interaction with some dense clouds here. There
are also clearly optical clouds with high velocity dispersion
associated with the jet--cloud impact \citep{Mundell03}. The radio
knots are near to the location of strong [FeII] emission \citep{SB09},
falling between the peaks of the [FeII] emission.

Altogether, the morphology suggests a scenario where the radio ejecta
runs into a denser cloud in the inhomogeneous ISM and results in
locally enhanced shock heating \citep[e.g.,][]{Capetti99}.  This
jet--cloud interaction has been suggested to explain the enhanced
[FeII] emission, which could be due to the iron unlocked from grains
by the shocks \citep{SB09}.  The NeIX emission, which traces the
higher ionization gas, becomes more prominent with the extra
collisional ionization from the jet--cloud interaction.
\citet{Armentrout07} also noted that the poor fit to the OVII line
profile in the grating spectra could be due to presence of a
non-photoionization component.  This possibility is further supported
by our spectral fits, which suggest that the NeIX/OVII enhancements
have a thermal origin, requiring the presence of a collisionally
ionized plasma (the $kT=0.58\pm 0.05$~keV component) in addition to a
photoionized nuclear outflow \citep{Yang01,Wang09_HRC}.

The emission measure of the $APEC$ component allows us to estimate the
electron density $n_e$ ($\approx n_H$) and the thermal pressure
$p_{th}$ ($\sim 2n_ekT$).  The emitting volume ($V$) of the hot gas,
assuming that the depth along the line of sight is comparable to the
other dimensions, is $5.6\times 10^{60}$ cm$^{3}$ for the high
NeIX/OVII ratio hot spots.  The filling factor $\eta$ is assumed to be
100\% and both of $n_e$ and $p_{th}$ have a weak dependence on $\eta$
($p \propto \eta^{-1/2}$).  There is $\sim 10\%$ uncertainty in $kT$,
which is propagated to the estimated pressure and energy.  For the
$kT=0.58$~keV emission, we derived an average density $n_e=0.06$
cm$^{-3}$, a hot gas pressure $p_{th}=6.8\times 10^{-10}$ dyne
cm$^{-2}$, a thermal energy content of $E_{th}=4\times 10^{51}$ ergs,
and a cooling time of $\tau_c=E_{th}/L_{APEC}\sim 10^5$ yr.  The
estimated internal pressure of the radio jet based on the synchrotron
minimum energy assumption ranges between $10^{-7}$ and $10^{-9}$ dyne
cm$^{-2}$ (Pedlar et al. 1993), to be confined by the hot gas pressure
of $10^{-9}$ dyne cm$^{-2}$, which further supports our collisional
ionization scenario.

Assuming $T=10^4$~K for the photoionized clouds and $n_e=220$
cm$^{-3}$ \citep{Penston90}, the thermal pressure from the
photoionized gas is $p_{ph}=3\times 10^{-10}$ dyne cm$^{-2}$, which is
comparable to the thermal pressure of the hot ISM, implying a possible
pressure equilibrium between the collisionally ionized hot gas and the
photoionized line-emitting cool clouds.

We further estimate the age of the interaction from the approximate
crossing time of the $0.5 \arcsec$ region \citep[dimension of the
  radio knot;][]{Pedlar93} with a characteristic velocity of
$c_{s}=200$ km s$^{-1}$, the local sound speed.  This is approximately
the thermal velocity for the 0.58~keV gas ($v_{th}=240$ km s$^{-1}$),
and also at the order of velocity dispersion of emission line gas seen
in \citet{SB10}.  Assuming the strong shock jump conditions, a crude
estimate of the shock velocity $v_{sh}$ can be obtained using $T
\approx 1.5\times 10^5$~K $(v_{sh}/100$~km s$^{-1})^2$ \citep{Raga02}.
For the $T\sim 6.7\times 10^6$~K X-ray-emitting gas, a $v_{sh}$ of
$\sim$700 km s$^{-1}$ relative to the downstream material is required.
This appears consistent with a supersonic but relatively slow radio
jet, which is constrained to be sub-relativistic from radio proper
motion estimates \citep[e.g., $v_{jet}< 12000$~km
  s$^{-1}$;][]{Ulvestad05}.  Thus we obtain a characteristic timescale
of $t_{cross}\sim 10^5$ yr.

If the jet--cloud interaction converts kinematic energy into heating
of the hot gas, a lower limit can be placed on the kinematic
luminosity of the jet, $L_{K.E.} \ga E_{th}/t_{cross}=1\times 10^{39}$
erg s$^{-1}$.  It is interesting to compare this observed $L_{K.E.}$
to the jet power directly inferred from the synchrotron emission.
Using the $P_{jet}$--$P_{radio}$ scaling relation (Equation 1) in
\citet{Cavagnolo10} and a total radio luminosity at 1.4 GHz
\citep[$\nu L_{\nu}\approx 4\times 10^{37}$ erg
  s$^{-1}$;][]{Mundell99}, we find $P_{jet}\sim 10^{42}$ erg s$^{-1}$,
suggesting that $\ga 0.1\%$ of the jet power is deposited in the ISM.
However, recalling that the $P_{jet}$--$P_{radio}$ relation was
derived from a sample of radio loud galaxies with X-ray cavities (see
\citealt{Cavagnolo10} for review on the $P_{jet}$--$P_{radio}$ scaling
relation) while here NGC 4151 is of low radio power, we also estimate
directly the jet energy flux using the pressure in the knots
\citep{Pedlar93} based on the minimal energy assumption, giving a
similar $P_{jet}=5\times 10^{42}$ erg s$^{-1}$.

We further note that, the crossing time $t_{cross}$ is comparable to
the cooling time $\tau_c$, implying that the hot gas is prominent only
locally (close to the radio knots).  Indeed, along the path of the
radio jet, the clouds that are spatially close to the jet impact spots
(e.g., clouds \#3,4 bracketing knot C2; \#9,10 for knot C5) show
[OIII]/X-ray ratios indistinguishable from others
(Figure~\ref{ratio}).  If these clouds are physically related to the
ISM interacting with the jet, instead of being projected close to the
knots in our line-of-sight, their X-ray emission shows little evidence
for excess shock heating in addition to nuclear photoionization.  This
localized heating is perhaps related to the highly collimated radio
ejecta in NGC 4151 \citep{Mundell03}, in contrast to the more expanded
lobe-like radio outflow in Mrk 3 \citep{Capetti99}, where the NLR gas
appears broadly impacted.  We plan to investigate in future work
whether the [OIII]/X-ray ratio, such as $\la$4 measured here and seen
in Mrk 3 and NGC 1386 \citep{Bianchi06}, together with other line
ratios (e.g., [FeII]/[PII]; \citealt{Oliva01}) could be useful
diagnostics for strong jet--ISM interaction in Seyfert galaxies.

In terms of energetics, we summarize our findings in NGC 4151: (a) The
jet power estimated from the radio power is $\sim 10^{42}$ erg
s$^{-1}$, a few percent of the current AGN energy output
\citep[$L_{bol}= 7\times 10^{43}$ erg s$^{-1}$;][]{Kaspi05}, which is
close to the canonical $\sim$5\% of the radiant energy adopted in
theoretical models for efficient AGN feedback
\citep[e.g.,][]{DiMatteo05,Hopkins06}; (b) $\ga 0.1\%$ of the jet
power is observed to have been deposited into the host ISM in the
nuclear region through the interaction between the radio jet and the
dense medium; (c) The shock heating due to the jet--cloud interaction
appears localized to the impact spots, and most clouds are consistent
with being photoionized by the nucleus.

\subsection{Extent and Origin of the Fe K$\alpha$ Line}

\citet{Ogle00} claimed that the narrow iron line emission was
spatially resolved in the {\em Chandra} HETG observation and $65\% \pm
9\%$ of the FeI K$\alpha$ emission comes from the ENLR at distance up
to 6\arcsec extent ($\sim$400 pc across).  This is intriguing, since
the putative pc-scale torus or more compact broad-line region
\citep[][and references therein]{Liu10} are expected to be the primary
location of narrow FeI K$\alpha$ emission.  \citet{Schurch03}
cautioned that due to the sensitivity of the short grating observation
in \citet{Ogle00}, the 1--3 arcsec region off the peak of
cross-dispersion profile contains few photons. Therefore the detection
of spatially extended iron line emission cannot be significant.  A
high signal-to-noise XMM-Newton spectrum of the neutral Fe K$\alpha$ suggests
that all line flux originates in a nearly Compton-thick torus that is
not resolvable at current resolution \citep{Schurch02,Schurch03}.

Our results show that the extended Fe K$\alpha$ emission, if present,
is only at $\sim 5\%$ level of the observed Fe K$\alpha$ emission,
when the contribution from the PSF wings is taken into account.  This
is consistent with our constraint on the extent of the Si fluorescent
emission and supports the conclusions of \citet{Schurch03}, while it
is in strong disagreement with the $65\%$ reported in \citet{Ogle00}.
We note that in \citet{Ogle00} the PSF is represented as a narrow
Gaussian with $FWHM=0.9\arcsec$, when the {\em Chandra} mission was
new.  The MARX model is now much more advanced so that a more reliable
estimate becomes possible.  At 6.4 keV, the encircled energy fraction
of a point source is only $63.7\%$ (see $Chandra$ Proposers'
Observatory Guide\footnote{\url{http://cxc.harvard.edu/proposer/POG/}}
Chapter 4 Table 4.2) in such a PSF core.  Hence a significant fraction
of the nuclear emission could have been measured in Ogle et al. as
spatially extended Fe K$\alpha$ emission.

\section{Conclusions}

In this paper we present spectral analysis and emission line images
from deep {\em Chandra} observation of NGC 4151, aiming to resolve and
characterise the X-ray emission in the inner $\sim$130 pc-radius
nuclear region.  The findings are summarized as follows:

\begin{enumerate}

\item We have obtained high spatial resolution X-ray narrow-band
  images of OVII, OVIII, and NeIX line emission, which are blended at
  the ACIS spectral resolution.  The images show extended structures
  that are spatially correlated with the radio outflow and optical
  [OIII] emission.

\item We find strong evidence for jet--ISM interaction, including
  morphological correspondences with regions of X-ray enhancement,
  peaks of NIR [FeII] emission, and optical clouds.  This is further
  strengthened by the presence of a $kT=0.58$~keV collisionally
  ionized component in the spectral fitting of the hot spots, and by
  the excess of X-ray emission in addition to nuclear photoionization
  as indicated by a low [OIII]/X-ray ratio. We find a possible
  pressure equilibrium between the collisionally ionized hot gas and
  the photoionized cool clouds.  The estimated velocity of the shocks
  from the jet--cloud impact is $\sim$700 km s$^{-1}$.

\item We estimate that the jet power in NGC 4151 is close to a few
  percent of the current AGN energy output ($L_{bol}\sim 7\times
  10^{43}$ erg s$^{-1}$).  The derived thermal energy in the hot gas
  suggests that $\ga 0.1\%$ of the jet power is deposited into the
  host ISM in the nuclear region through the interaction between the
  radio-jet and the dense medium.  The [OIII]/X-ray ratios of NLR
  clouds bracketing the radio knots show little deviation from other
  photoionized clouds, indicating a localized impact on the ISM by the
  highly collimated jet.

\item We investigate the spatial extent of the fluorescent features,
  including the Fe K$\alpha$ emission and the Si K$\alpha$ emission.
  Our results show that both are dominated by point-like emission,
  consistent with an origin of unresolved inner structure such as a
  torus.  The extended Fe K$\alpha$ emission is $\la 5\%$ of the
  observed Fe K$\alpha$ emission, which is in strong disagreement with
  the $65\%$ reported in \citet{Ogle00}.

\end{enumerate}

\acknowledgments

We thank the anonymous referee for helpful suggestions. This work is
supported by NASA grant GO8-9101X (PI: Fabbiano) and grant GO1-12009X
(PI: Wang).  We acknowledge support from the CXC, which is operated by
the Smithsonian Astrophysical Observatory (SAO) for and on behalf of
NASA under Contract NAS8-03060.  CGM acknowledges financial support
from the Royal Society and Research Councils U.K.  J. W. thanks
G. Ferland, T. Kallman, S. Bianchi, A. Marinucci, and S. Chakravorty
for extensive advices on photoionization modeling, P. Nulsen for jet
power discussion, T. Storchi-Bergmann and R. Riffel for providing the
Gemini NIFS maps.  This research has made use of data obtained from
the {\em Chandra} Data Archive, and software provided by the CXC in
the application packages CIAO and Sherpa.

{\it Facilities:} \facility{CXO (HRC, ACIS)}




\clearpage

\clearpage

\clearpage

\clearpage

\clearpage

\pagestyle{empty}
\begin{deluxetable}{cccccccccccc}
\tabletypesize{\scriptsize}
\tablewidth{0pt}
\rotate
\tablecaption{Spectral Fits to the NeIX/OVII Enhancements and Surrounding Region \label{tab:cloudy}}
\tablecolumns{12} 
\tablehead{\colhead{Region} & \colhead{Counts}  & \colhead{$\log U_1$} & \colhead{$\log N_{H,1}$}& \colhead{$Norm1$}  & \colhead{$\log U_2$} & \colhead{$\log N_{H,2}$} & \colhead{$Norm2$} & \colhead{$kT$} & \colhead{$Norm3$} & \colhead{Nuclear PSF\tablenotemark{a}} & \colhead{$F_{0.3-2keV}$\tablenotemark{a}}\\ 
\colhead{} & \colhead{[0.3--2 keV]} & \colhead{} & \colhead{[cm$^{-2}$]} & \colhead{} & \colhead{} & \colhead{[cm$^{-2}$]}  & \colhead{} & \colhead{[keV]}  & \colhead{}  & \colhead{} & \colhead{[erg s$^{-1}$ cm$^{-2}$ arcsec$^{-2}$]}}

\startdata
High-ratio & 4585 & $-0.01\pm 0.03$  & 20.5 & $8.0\times 10^{-16}$  & $1.9\pm 0.2$ & 22.5 & $1.5\times 10^{-18}$  & $0.58\pm 0.05$ & $1.8\times 10^{-5}$ & $1455\pm 35$ & $8.9\times 10^{-14}$ \\
Low-ratio  &  1317 & $-0.01\pm 0.03$ & 20.5 & $8.2\times 10^{-16}$  & $1.9\pm 0.2$ &  22.5 & $6.3\times 10^{-19}$ & $0.58\pm 0.05$ & $4.5\times 10^{-6}$& $530\pm 23$ & $6.8\times 10^{-14}$ \\
\enddata

\tablenotetext{a}{The absorption column $N_H$ is fixed at $2\times
  10^{20}$ cm$^{-2}$ \citep{Murphy96}, the Galactic column towards NGC
  4151; $\chi^2$ / d.o.f.= 146/142 for the best fit presented here.}

\tablenotetext{b}{Expected counts from simulation of PSF scattered
  nuclear emission in the 0.3--2 keV band.}

\tablenotetext{c}{Soft X-ray surface brightness of the extended
  emission after nuclear emission removed.}

\end{deluxetable}

\begin{deluxetable}{cccccccc}
\rotate
\tabletypesize{\scriptsize}
\tablecaption{Measured X-ray and [OIII] Fluxes\label{flux}}
\tablewidth{0pt}
\tablehead{
\colhead{\begin{tabular}{c}
Cloud\\
label\\
\end{tabular}} &
\colhead{\begin{tabular}{c}
Distance\\
to Nuc. ($\arcsec$)\\
\end{tabular}} &
\colhead{\begin{tabular}{c}
Distance\\
to Nuc. (pc)\\
\end{tabular}} &
\colhead{\begin{tabular}{c}
[OIII] flux\\
($10^{-13}$erg s$^{-1}$ cm$^{-2}$)\\
\end{tabular}} &
\colhead{\begin{tabular}{c}
Net Counts\\
(0.3-2 keV)\\
\end{tabular}} &
\colhead{\begin{tabular}{c}
0.5-2 keV Flux\tablenotemark{a}\\
($10^{-14}$erg s$^{-1}$ cm$^{-2}$)\\
\end{tabular}} &
\colhead{\begin{tabular}{c}
[OIII]/soft X\\
(ACIS) \\
\end{tabular}} &
\colhead{\begin{tabular}{c}
[OIII]/soft X\\
(HRC) \\
\end{tabular}}
}
\startdata
1 & 1.76 & 114 & 1.1 & 369$\pm$24 & 1.5 & 7.4 & 12\\
2 & 1.44 & 93.6 & 1.9 & 688$\pm$29 & 2.7 & 6.9 & 12\\
3 & 1.01 & 65.6 &2.2 & 581$\pm$38 & 2.3 & 9.4 & 12\\
4 & 0.78 & 50.7 &1.3 & 470$\pm$39 & 1.9 & 6.9 & 7\\  
5 & 1.0 & 65.0 &1.5 & 168$\pm$29 & 0.7 & 22.2 & 75\\
6 & 0.6 & 39. &1.1 & 99$\pm$44 & 0.4 & 27.8 & 110\\
7 & 0.4 & 26. &3.4 & 1320$\pm$96 & 5.3 & 6.4 & 5\\
8 & 0.4 & 26. &3.3 & 560$\pm$100 & 2.2 & 14.7 & 13\\
9 & 0.85 & 55.2 &2.4 & 694$\pm$47 & 2.8 & 8.6 & 3\\
10 & 0.89 & 57.8 &1.6 & 266$\pm$35 & 1.1 & 15.0 & 8\\
11 & 2.15 & 139.7 &0.6 & 264$\pm$17 & 1.1 & 5.7 & 15\\
\hline
C2\tablenotemark{b} & 0.9 & 58.5 & 0.8 & 1180$\pm$44 & 4.7 & 1.7 & 2\\
C5\tablenotemark{b} & 0.93 & 60.4 & 1.4 & 1052$\pm$43 & 4.2 & 3.3 & 3\\
\enddata

\tablenotetext{a}{An average conversion factor $7.2\times 10^{-12}$ erg s$^{-1}$ cm$^{-2}$ (counts s$^{-1}$)$^{-1}$ is derived from spectral models in Table~1 to estimate flux from net count rate.}

\tablenotetext{b}{C2 and C5 are the radio knots in Mundell et al.\ (1995), corresponding to the locations of high NeIX/OVII ratio hot spots to the east and west of the nucleus, respectively.}
\end{deluxetable}

\end{document}